\documentstyle[sprocl,epsfig]{article}

\def\Journal#1#2#3#4{{#1} {\bf #2}, #3 (#4)}
\def\AP{\em Ann. Phys.}

\def\NPA{{\em Nucl. Phys.} A}
\def\NPB{{\em Nucl. Phys.} B}

\def\PLB{{\em Phys. Lett.} B}

\def\PRL{\em Phys. Rev. Lett.}

\def\PREP{\em Phys. Rep.}
\def\PRD{{\em Phys. Rev.} D}
\def\PRC{{\em Phys. Rev.} C}

\def\ZP{\em Z. Phys.}
\def\JETP{\em JETP}

\begin{document}

\title{
MASSES OF HADRONS IN NUCLEI\footnote{Plenary talk at BARYONS~98, 
Bonn, Germany, September 1998} }

\author{B. FRIMAN}
\address{GSI, Plankstr. 1, D-64291 Darmstadt, Germany and\\
Inst. f\"ur Kernphysik, TU Darmstadt, D-64289 Darmstadt, Germany}

\author{M. LUTZ}
\address{GSI, Plankstr. 1, D-64291 Darmstadt, Germany}

\author{G. WOLF}
\address{KFKI, RMKI, H-1525 Budapest POB. 49, Hungary}


\maketitle

\abstracts{
We emphasize the central role played by the spectral function in
the description of hadrons in matter and discuss the applicability
of the quasiparticle concept to the propagation of hadrons in dense
nuclear matter. Theoretical and experimental results relevant for
the in medium properties of vector mesons and kaons are briefly
reviewed. We also present novel results for the $\rho$ and $\omega$
spectral functions in nuclear matter, deduced from a coupled
channel analysis of pion-nucleon scattering data.}

\section{Introduction}

The in-medium properties of hadrons is a topic of high current
interest in the hadron and nuclear physics community. During the
last few years the discussion has focused on kaons and light vector
mesons because relevant experimental data has been available for
these mesons. Indeed, in relativistic nucleus-nucleus collisions
one finds enhancements in the spectrum of low-mass lepton
pairs~\cite{CERES,Helios3} and in the $K^-$
multiplicity~\cite{kaos}. Presently several distinct theoretical
interpretations of these findings exist. However, in almost all
models that successfully reproduce the data, medium effects of some
kind are invoked. Consequently, it seems likely that these data
reflect nontrivial properties of dense hadron matter.

We begin by making some general remarks on quasiparticles in
many-body systems, and then briefly recall the fairly well
established properties of nucleons in nuclei as well as the
expected characteristics of kaons and anti-kaons in nuclear matter.
In the subsequent part we discuss theoretical models for the
properties of vector mesons in matter and the experimental data on
lepton-pair production in relativistic nucleus-nucleus collisions.
Finally, we present a coupled channel approach to meson-nucleon
scattering. In this approach, we fit the available meson-nucleon
scattering data and extract the vector-meson--nucleon scattering
amplitudes. With these amplitudes we then construct the
vector-meson self energies in nuclear matter to leading order in
density.

\section{Quasiparticles}

The quasiparticle concept, which was introduced by Landau in his
theory of Fermi liquids~\cite{landau}, has been very successfully
applied to a rich variety of phenomena in liquid $^{3}$He and other
low-temperature Fermi system~\cite{bp}. Landau's theory was extended
to nuclear physics by Migdal~\cite{migdal}. In Fermi liquid theory
there is a one-to-one correspondence between the quasiparticle
excitations of the interacting system and the single-particle
excitations of the non-interacting system. In other words, a nucleon
added to a nucleus turns into a quasiparticle (quasi-nucleon) when the
interaction is turned on. In the interacting many-particle system, the
single-particle strength is spread over many eigenstates of the
Hamiltonian, i.e., the single-particle state is mixed with complicated
many-particle states. Thus, the mass (or some other property) of a
particle in an interacting many-body system, e.g. a hadron in a
nucleus, is a priori {\it not} a well defined concept. Only if part of
the single-particle strength is concentrated in a set of eigenstates
that are close in energy, while the rest of the strength is spread
over an almost uniform background, one can identify a unique
quasiparticle. In that case one equates the in-medium properties of
the particle with those of the corresponding quasiparticle.

The properties of the quasiparticle (mass, magnetic moment etc.)
generally differ from those of the free particle. Moreover, because
its strength is spread over several energy eigenstates, a
quasiparticle normally has a width $\Gamma_{QP}$ and consequently a
finite lifetime $\tau_{QP} = 1/\Gamma_{QP}$, also when the
corresponding particle is stable in vacuum. Thus, a quasiparticle
typically corresponds to a peak with a finite width in the spectral
function. We note that in mean-field approximations the
quasiparticle width is neglected, i.e., the existence of a well
defined quasiparticle is postulated. In general, the quasiparticle
concept is useful for describing the response of the many-body
system to probes, whose characteristic time scale is shorter than
the quasiparticle lifetime. Whether this is the case for a given
system, must be checked either by experiment or by calculation.

\begin{figure}[t]
\center{\epsfig{file=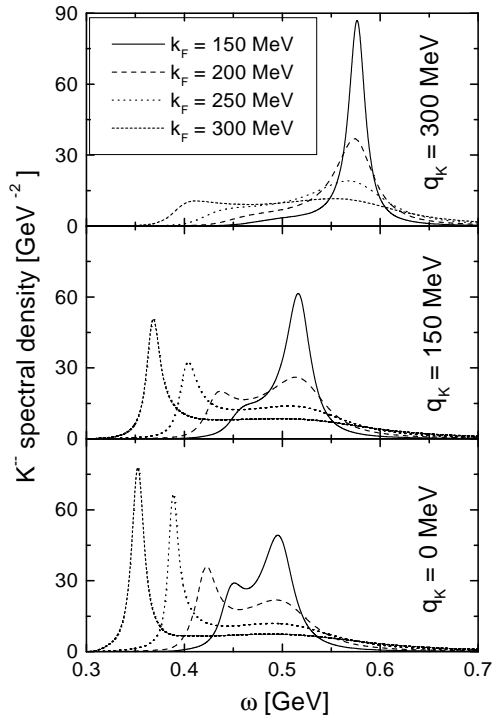,height=100mm}}
\caption{\label{kaons} The $K^-$ spectral function in nuclear
matter at several densities and momenta from ref.~$^9$. The
saturation density of nuclear matter, $\rho_0=0.17\mbox{~fm}^{-3}$
corresponds to $k_F=268$ MeV.}
\end{figure}

If the spectral function does not show a clear structure that can
be identified with a quasiparticle peak, it obviously makes no
sense to discuss the in-medium mass of the corresponding particle.
In such a case the response of the many-body system is best
described by means of the full spectral function. Detailed
calculations show that this may in fact be the case for rho
mesons~\cite{peters,mosel} and negative kaons~\cite{lutz} in
nuclear matter. In fig.~\ref{kaons} we show the $K^-$ spectral
function in nuclear matter of ref.~\cite{lutz}. Already at moderate
densities and certain values of the 3-momentum, the kaon spectral
density shows structures that are not described by a quasiparticle
ansatz.

\section{Hadrons in matter}

The quasiparticles best known to nuclear physicist are nucleons in
nuclei. Since the quasiparticles in Fermi systems are subject to
the Pauli principle their widths vanish at the Fermi surface. Thus,
a quasi-nucleon close to the Fermi surface has a long lifetime and
consequently appears as an independent particle. This is
essentially the reason for the success of independent particle
models, like the shell model, in describing the structure of
nuclei. The properties of the quasi-nucleons differ from those of
free nucleons. An important property is the nucleon effective mass
$m^\star$, defined by
\begin{equation}
\frac{p}{m^\star} = \frac{\mbox{d}\varepsilon_p}{\mbox{d} p},
\end{equation}
which characterizes the non-locality of the average nuclear
potential. One finds~\cite{bgg,mahaux} that the nucleon effective
mass close to the Fermi surface is almost equal to the free mass,
while 10-20 MeV above or below the Fermi energy it is approximately
0.7 m. Empirical constraints on the effective mass are obtained
from the single-particle level density and phenomenological optical
potentials for proton-nucleus scattering.

The in-medium properties of kaons have recently been the subject of
intense discussions. In central heavy-ion collisions at
sub-threshold energies, the KaoS collaboration at GSI~\cite{kaos}
finds a substantial enhancement the of $K^-$ yield per participant
compared to proton-proton collisions. This effect can be
interpreted in terms of a strong attractive medium modification of
the $K^-$-spectrum, which facilitates the production of anti-kaons
in the medium~\cite{li,cassing}. This scenario is consistent with
calculations of kaon properties in nuclear
matter~\cite{chlee,BRkaon,waas,lutz}, which predict a weakly
repulsive mass shift for $K^+$ mesons and a very attractive, energy
dependent potential for the $K^-$ (see fig.~1). The predictions for kaon
properties in nuclear matter are fairly robust at moderate
densities since they are constrained by kaon-nucleon scattering
data. We note, however, that the $K^-$-spectral density is subject
to uncertainties  since it probes the $K^-$-nucleon scattering
process at sub-threshold kinematics which is not directly accessible
in scattering experiments. Consequently, any prediction of the
$K^-$-mass distribution is closely linked to the in-medium dynamics
of the $\Lambda (1405)$ resonance~\cite{koch,lutz},which plays a
crucial role in $K^-$-nucleon scattering~\cite{martin}.

The electromagnetic decay of vector mesons into $e^+e^-$ and
$\mu^+\mu^-$ pairs makes them particularly well suited for
exploring the conditions in dense and hot matter in nuclear
collisions. The lepton pairs provide virtually undistorted
information on the spectral distribution of the vector mesons in
the medium. The low mass enhancement of lepton pairs found in
ultra-relativistic nucleus-nucleus collisions by the CERES and
HELIOS-3 collaborations~\cite{CERES,Helios3} at CERN cannot be
interpreted in terms of ``standard'' models, where the properties
of the hadrons are not modified in the medium~\cite{KKL}.

Since the data show an enhancement compared to the results of such
calculations at invariant masses below the $\rho/\omega$ peak, a
natural interpretation of the enhancement could be that the
in-medium masses of the light vector mesons are strongly reduced in
dense/hot hadron matter. Indeed, Li, Ko and Brown find very good
agreement with the data in the universal scaling scenario, where
the $\rho$ and $\omega$ masses decrease with increasing baryon
density~\cite{likobrown}. This scenario is based on the universal
scaling approach of Brown and Rho~\cite{BR}, where the masses of
all hadrons, except for pseudo-scalar mesons, drop in proportion to
the quark condensate.

An alternative interpretation of the low-mass enhancement of lepton
pairs, based on effective hadronic models for the vector-meson self
energy in matter, has been explored by Rapp, Chanfray and
Wambach~\cite{RCW}. In this approach, the starting point is an
effective hadron Lagrangian, whose parameters i\-deal\-ly are
determined by hadronic interactions in vacuum. The vector-meson
properties in matter are then computed by evaluating the
corresponding self energy using standard many-body techniques.
Several groups have done calculations along these lines, with
qualitatively similar results. One finds a strong enhancement of
the in-medium width of the $\rho$ meson due to the strong
interaction of the pion cloud with the surrounding
nucleons~\cite{AK,CS,HFN,KKW,peters}, as well as a momentum
dependence of the $\rho$-meson spectral function due to s-channel
baryon resonances~\cite{FP,peters}. Both the enhanced width and the
momentum dependence leads to a shift of $\rho$-meson strength to
smaller invariant masses.  When these effects are implemented in a
simulation of heavy-ion collisions, Rapp {\em et al.} find good
agreement with the data~\cite{RCW}.

In the following section we present a recent calculation in an
effective hadronic model. This calculation is a first attempt to
include, in a systematic fashion, constraints from meson-nucleon
scattering data in the energy regime relevant for the properties of
vector mesons in nuclear matter ($\sqrt{s} \simeq 1.7$ GeV).
Previous calculations were based on effective interactions adjusted
to low-energy data only. These necessarily involve extrapolations
over a wide range in energy, which introduce a strong model
dependence~\cite{bfseoul} and consequently should be avoided
whenever possible.

\section{Meson-nucleon scattering}

In this section we describe a relativistic and unitary coupled
channel approach to meson-nucleon scattering~\cite{LWF}. The
following channels are included: $\pi N$, $\rho N$, $\omega N$,
$\pi \Delta$ and $\eta N$. Our goal is to determine the
vector-meson--nucleon scattering amplitude since it determines the
self energy of a vector-meson in nuclear matter to leading order in
density.
\begin{figure}[t]
\center{\epsfig{file=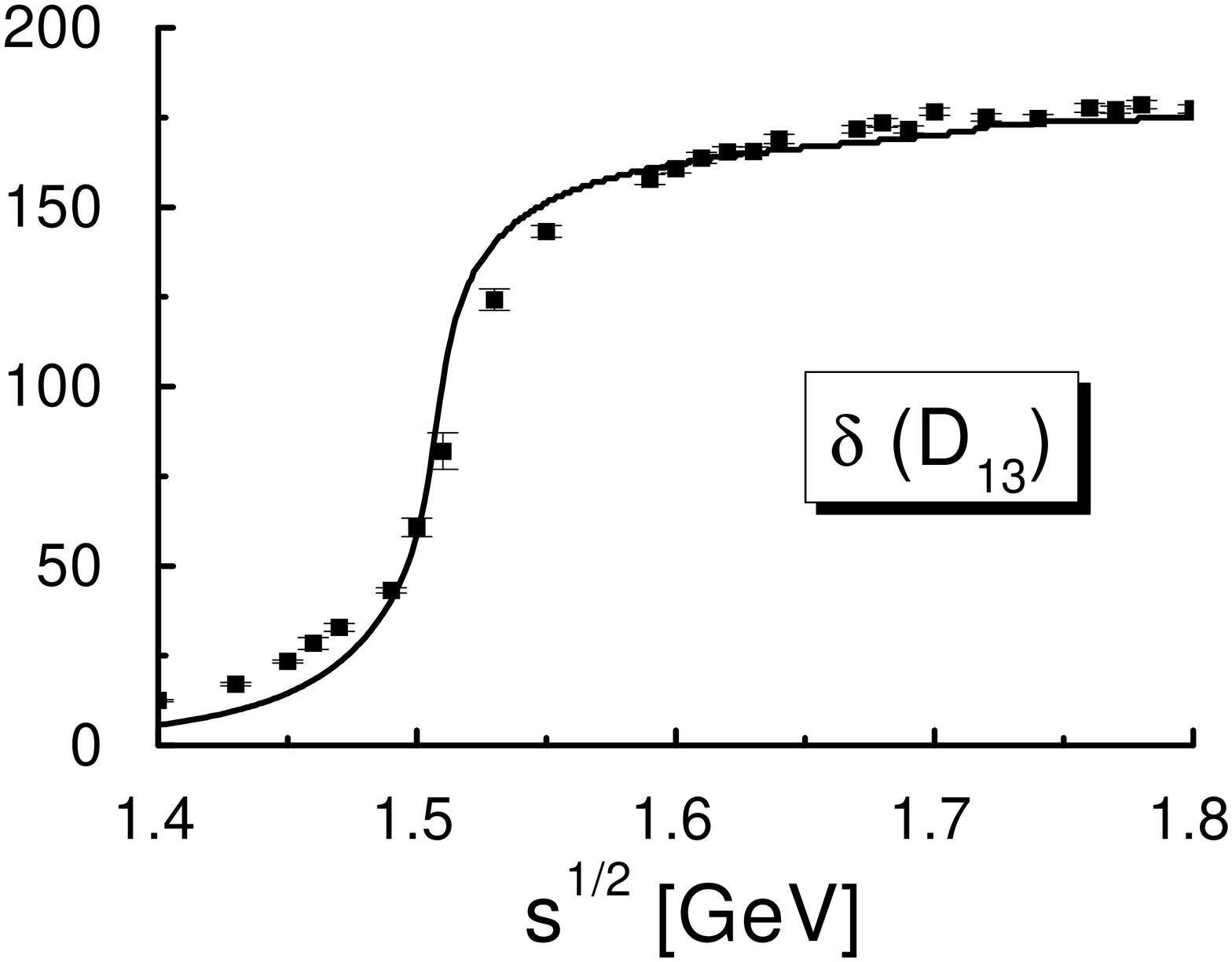,height=50mm}}
\center{\epsfig{file=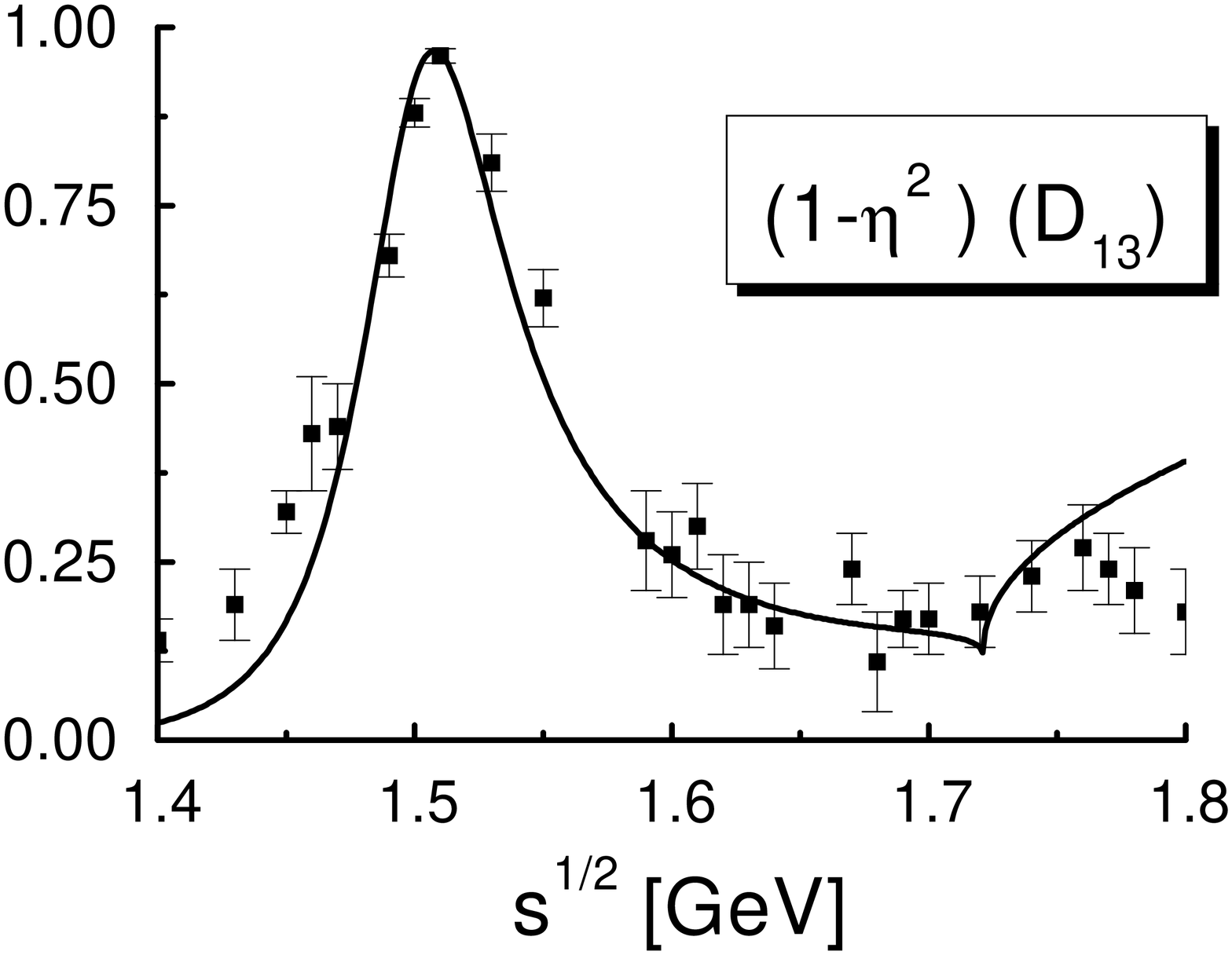,height=50mm}}
\caption{\label{D13-scatt} The $\pi N$ scattering phase shifts
and inelasticity in the $D_{13}$ channel. The line shows the best
fit, while the data points are those of the analysis of Arndt {\em
et al.}~$^{30}$.}
\end{figure}
In this work we focus on vector mesons with small or zero
3-momentum with respect to the nuclear medium. Therefore it is
sufficient to consider only s-wave scattering in the $\rho N$ and
$\omega N$ channels. This implies that in the $\pi N$ and
$\pi\Delta$ channels we need only s- and d-waves. In particular, we
consider the $S_{11}, S_{31}, D_{13}$ and $D_{33}$ partial waves of
$\pi N$ scattering. Furthermore, we consider the pion-induced
production of $\eta$, $\omega$ and $\rho$ mesons off nucleons.

We recall that at small densities the spectral function of a vector
meson in nuclear matter with energy $\omega $ and zero momentum probes
the vector-meson nucleon scattering process at $\sqrt{s} \sim
m_N+\omega $. In order to learn something about the momentum
dependence of the vector-meson self energy, vector-meson--nucleon
scattering also in higher partial waves would have to be considered.

In accordance with the ideas outlined above only data in the
relevant kinematical range will be used in the analysis. The
threshold for vector-meson production off a nucleon is at $\sqrt{s}
\simeq 1.7$ GeV. We fit the data in the energy range
$1.45$ GeV $\leq \sqrt{s} \leq 1.8$ GeV, using an effective
Lagrangian with local 4-point meson-meson--baryon-baryon
interactions. For details the reader is referred to
ref.~\cite{LWF}.

In fig.~\ref{D13-scatt} our fit to the $\pi N$ scattering data is
illustrated by the $D_{13}$ channel. In the remaining channels the
fit is of similar quality.
\begin{figure}[t]
\center{\epsfig{file=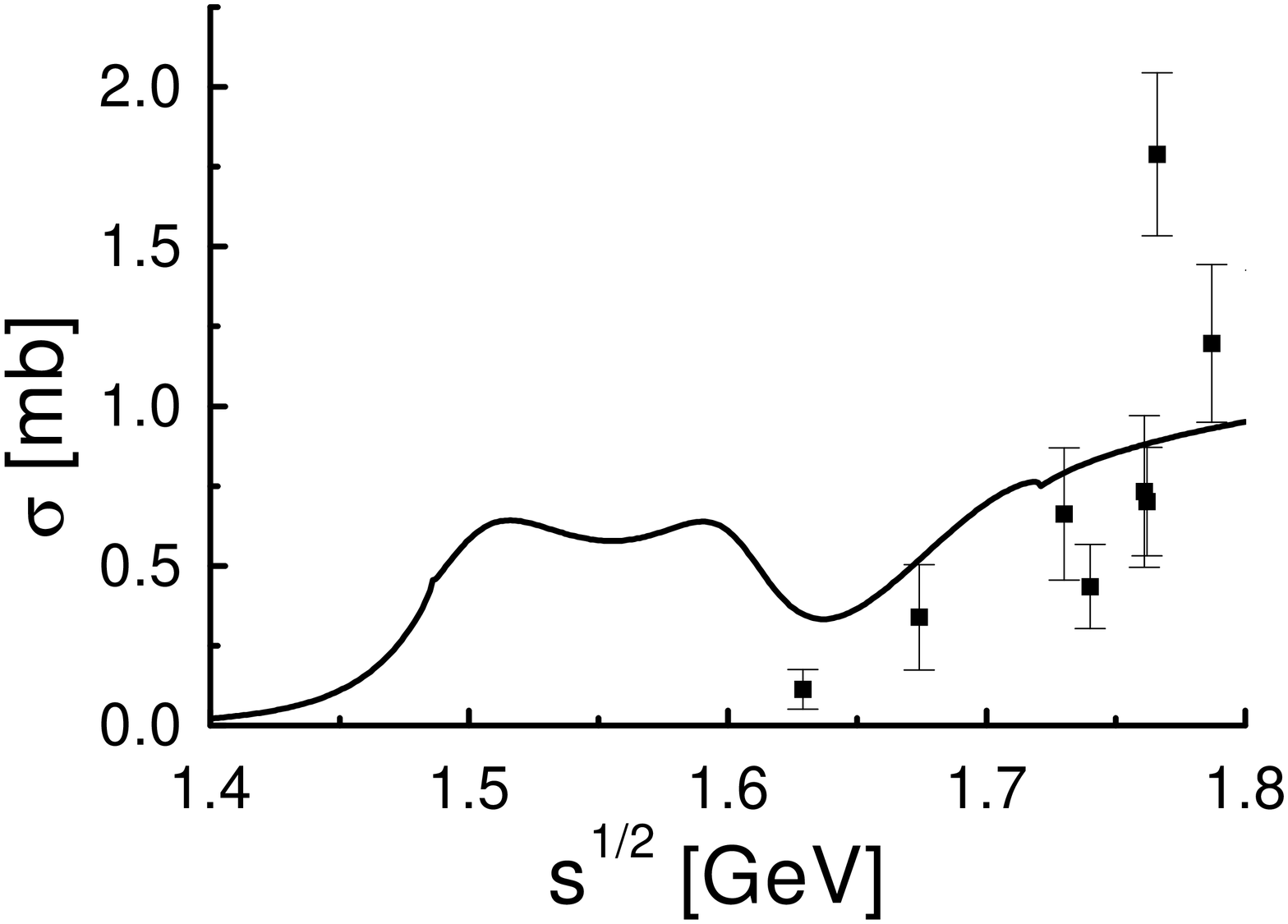,height=50mm}}
\center{\epsfig{file=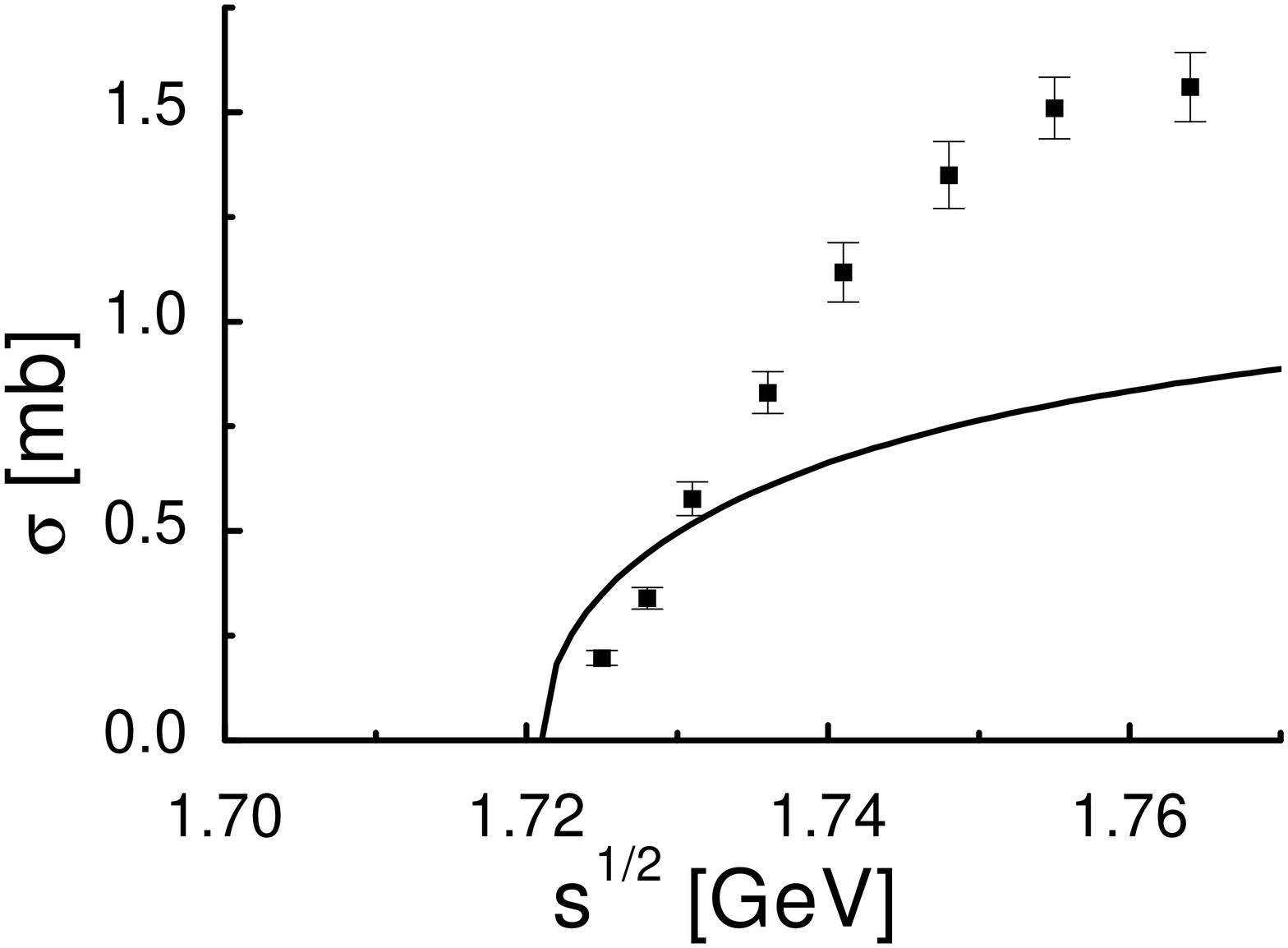,height=50mm}}
\caption{\label{rho-prod} The cross sections for the reactions
$\pi^- p \rightarrow \rho^0 n$ (top) and $\pi^- p \rightarrow \omega
n$ (bottom). The data are from ref.~$^{31,32}$ and 
ref.~$^{33}$, respectively.}
\end{figure}
Furthermore, in fig.~\ref{rho-prod} the cross sections for the
reactions $\pi^- p\rightarrow \rho^0 n$ and $\pi^- p\rightarrow
\omega n$ are shown. The agreement with the data is satisfactory
close to threshold, but the energy dependence is not reproduced.
This may be due to the coupling to channels, like the $K-\Sigma$
channel, as well as higher partial waves, not yet included in our
scheme.
\begin{figure}[t]
\center{\epsfig{file=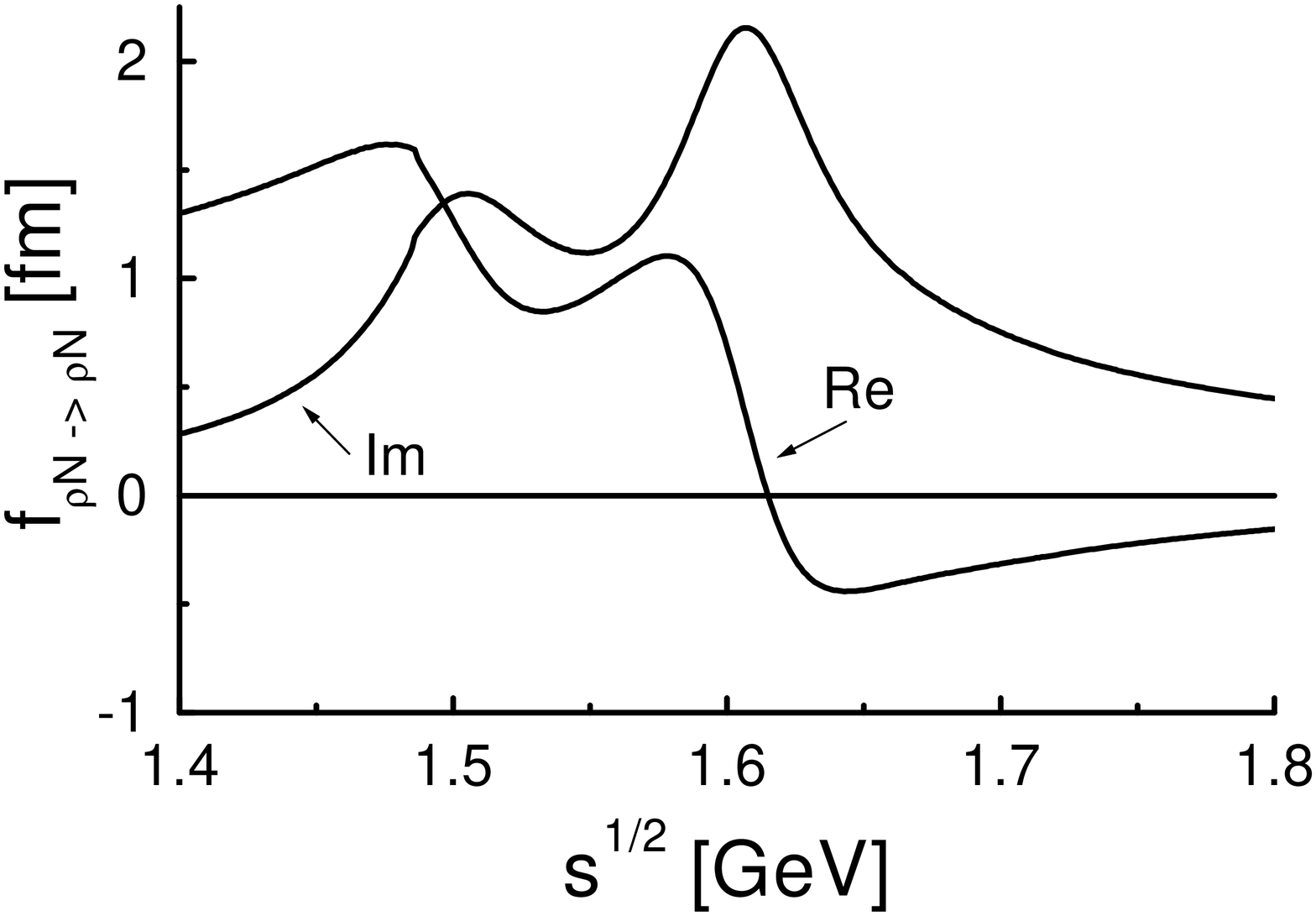,height=50mm}}
\center{\epsfig{file=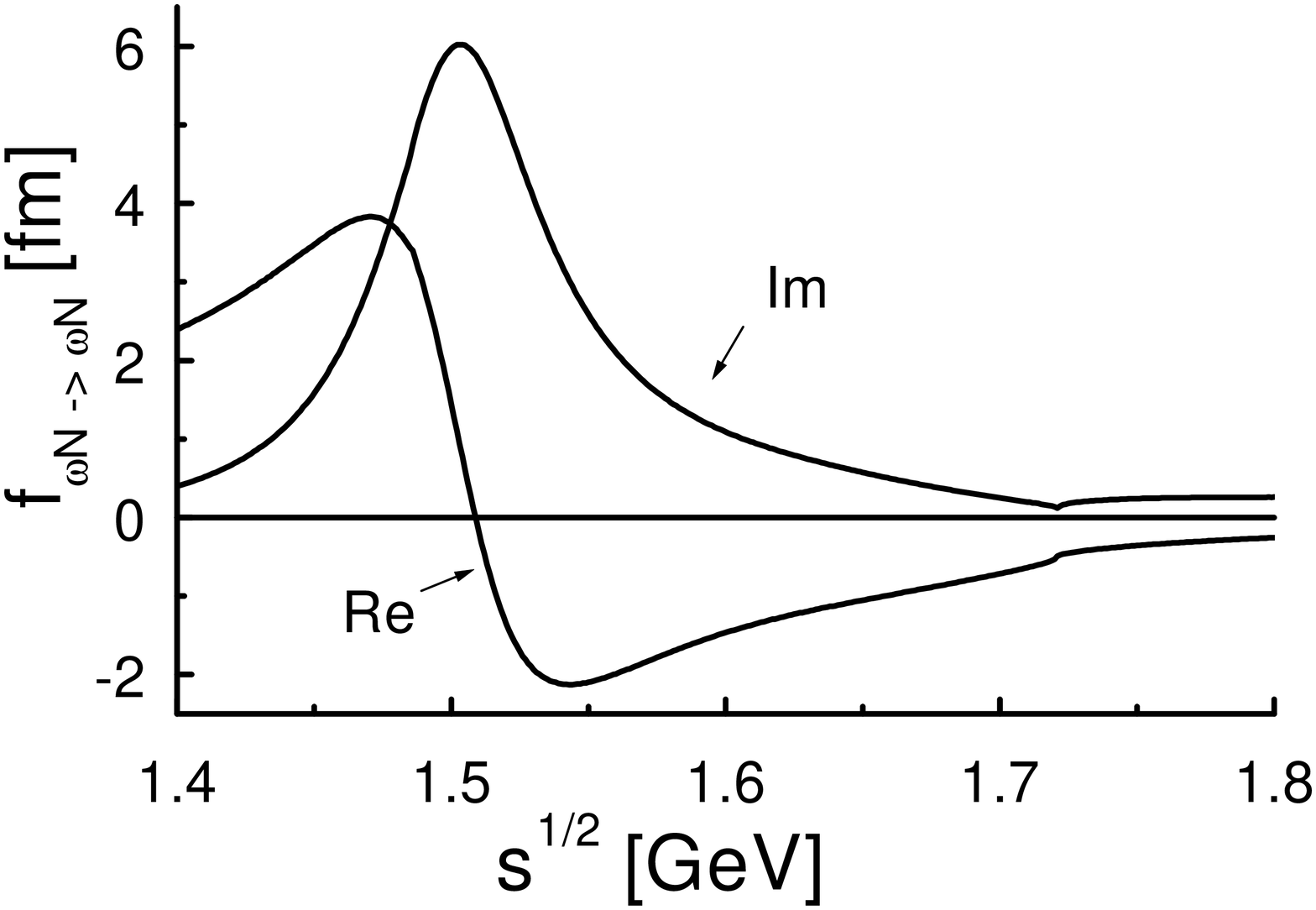,height=50mm}}
\caption{\label{amplitudes} The $\rho N$ and $\omega N$
scattering amplitudes, averaged over spin and isospin.}
\end{figure}

The bumps in the $\rho$-production cross section at $\sqrt{s}$
below 1.6 GeV are due to the coupling to resonances below the
threshold, like the $N^\star(1520)$. This indicates that these
resonances may play an important role in the $\rho$-nucleon
dynamics, in agreement with the results of Manley and
Saleski~\cite{manley}. However, the strength of the coupling is
uncertain, due to the ambiguity in the $\rho$-production cross
section close to threshold. We find that also the $\omega$ meson
couples strongly to these resonances.
The pion-induced $\eta$-production cross section (not shown) is
well described up to $\sqrt{s} \simeq 1.65$ GeV. At higher energies
presumably higher partial waves, presently not included in our
model, become important.

The resulting $\rho$- and $\omega$-nucleon scattering amplitudes
are shown in fig.~\ref{amplitudes}. The $\rho-N$ and $\omega-N$
scattering lengths, defined by $a_{VN}=f_{VN}(\sqrt{s}=m_N+m_V)$,
are $a_{\rho N}=$ (-0.3+0.7\,i) fm and $a_{\omega N}=$
(-0.5+0.1\,i) fm. To lowest order in density, this corresponds to
the following in-medium modifications of masses and widths at
nuclear matter density: $\Delta m_\rho
\simeq 30$ MeV, $\Delta m_\omega
\simeq 50$ MeV, $\Delta\Gamma_\rho \simeq 140$ MeV and $\Delta
\Gamma_\omega \simeq 20$ MeV. However, as we show in the next
section, the coupling of the vector mesons to baryon resonances
below threshold, which is reflected in the strong energy dependence
of the amplitudes, cannot be neglected.

\section{Vector mesons in nuclear matter}

\begin{figure}[t]
\center{\epsfig{file=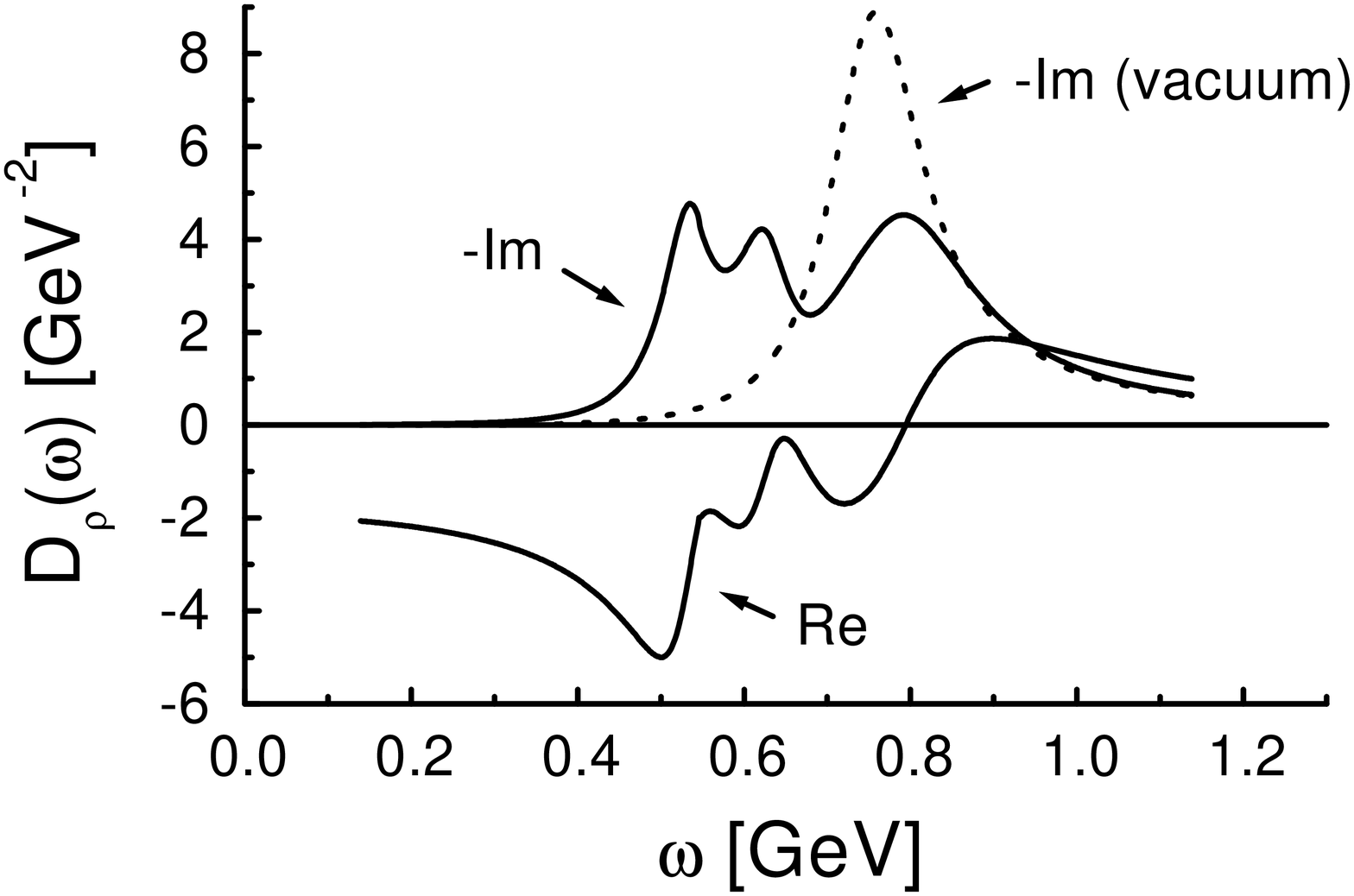,height=50mm}}
\center{\epsfig{file=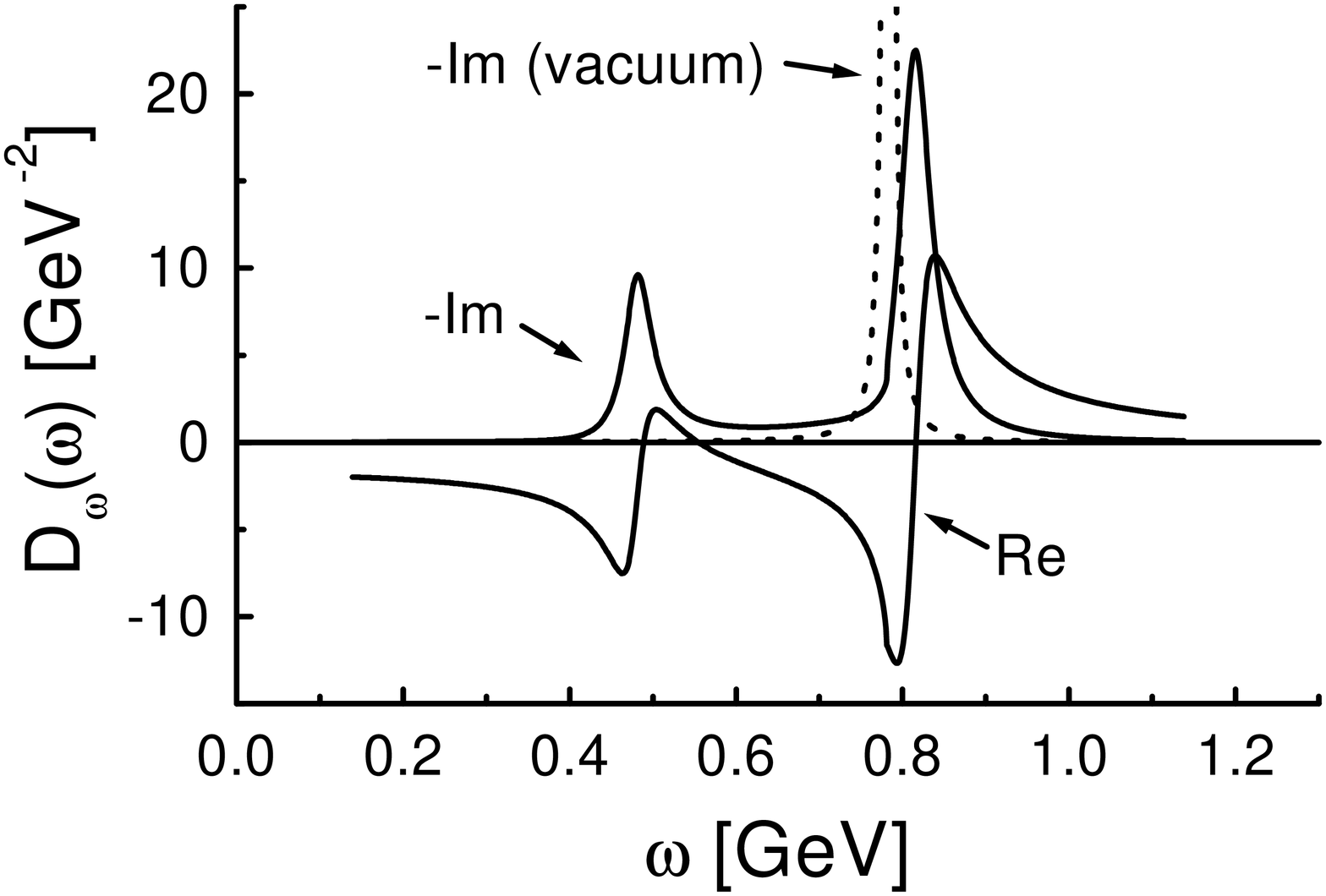,height=50mm}}
\caption{\label{propagators} Real and imaginary parts of the
$\rho$ and $\omega$ propagators in nuclear matter at $\rho_0$,
compared to the imaginary parts in vacuum.}
\end{figure}

In this section we present results for the in-medium propagators of the
$\rho$- and $\omega$-mesons at rest, obtained
with the scattering amplitudes presented in section 4, to leading
order in density. The low-density theorem states that the self
energy, $\Delta m_V^2(\omega )$, of a vector meson $V$ in nuclear
matter is given by~\cite{LDT}
\begin{equation}
\Delta m_V^2(\omega ) = -4\pi(1+\frac{\omega}{m_N})
f_{V N}\,(\sqrt{s}=m_N+\omega ) \rho_N + \dots,
\label{LDT}
\end{equation}
where $\omega$ is the energy of the vector meson, $m_N$ the nucleon
mass, $\rho_N$ the nucleon density and ${f}_{V N}$ denotes the $V
N$ s-wave scattering amplitude averaged over spin and isospin. In
fig.~\ref{propagators} we show the resulting propagators at the
saturation density of nuclear matter, $\rho_0 = 0.17
\mbox{~fm}^{-3}$. For the $\rho$ meson we note a strong enhancement
of the width, and a downward shift in energy, due to the mixing
with the baryon resonances at $\sqrt{s} = 1.5 - 1.6$ GeV. Thus, our
results lends support to the dynamical scenario discussed in
ref.~\cite{brown}. The center-of-gravity of the spectral function
is shifted down in energy by $\simeq 10$~\%.

The in-medium propagator of the $\omega$ meson exhibits two
distinct quasiparticles, an $\omega$ like mode, which is shifted up
somewhat in energy, and a resonance-hole like mode at low energies.
The low-lying mode carries about 20 \% on the energy-weighted sum
rule. Again, the center-of-gravity is shifted down by $\simeq
10$~\%. However, we stress that the structure of the in-medium
$\omega$ spectral function clearly cannot be characterized by this
number alone.

We expect that the results obtained with only the leading term in
the low-density expansion are qualitatively correct at normal
nuclear matter density. However, on a quantitative level, the
spectral functions may change when higher order terms in the
density expansion, induced e.g. by p-wave vector-meson--nucleon
interactions, are included.

\section{Conclusions}

We reviewed general aspects of quasiparticles in many-body systems
and emphasized the importance of the spectral function for the
discussion of hadron properties in matter. Furthermore, recent
theoretical and experimental developments relevant for the study of
kaons and vector mesons in a hadronic environment were presented.

We also reported on a relativistic and unitary, coupled
channel approach to meson-nucleon scattering. The parameters of the
effective interaction are determined by fitting elastic
pion-nucleon scattering and pion-induced meson production data in
the relevant energy regime. We obtain a model for the $\rho$- and
$\omega$-N scattering amplitudes, which allows us to compute the
vector-meson self energies in nuclear matter to leading order in
density. In this approach we avoid the extrapolation from
low-energy data, which is a weak point in previous calculations.

A prominent feature of the scattering amplitudes is the strong
coupling to baryon resonances below threshold. This leads to two
characteristic features of the vector-meson spectral functions,
namely repulsive scattering lengths and a spreading of the
vector-meson strength to states at low energy. The latter is, as
discussed above, qualitatively what seems to be required by the
heavy-ion data. Clearly, complementary experiments with e.g. photon
and pion induced vector meson production off nuclei would be
extremely useful for exploring the in-medium properties of these
mesons in more detail.


\end{document}